\documentclass{sig-alternate}

\begin{document}

\title{Deaf, Hard of Hearing, and Hearing perspectives on using Automatic Speech Recognition in Conversation}

\CopyrightYear{2017} 
\setcopyright{rightsretained} 
\conferenceinfo{ASSETS '17}{October 29 - November 1, Baltimore, MD, USA} 
\isbn{}
\doi{}

\numberofauthors{1}
\author{
\alignauthor
		Abraham Glasser, Kesavan Kushalnagar, and Raja Kushalnagar\\
       \affaddr{Rochester Institute of Technology}\\
       \affaddr{1 Lomb Memorial Drive}\\
       \affaddr{Rochester, NY 14623}\\
       \email{\{atg2036,krk4565, reuami\}@rit.edu}
}

\maketitle

\abstract{
In this experience report, we describe the accessibility challenges that deaf,
hard of hearing and hearing participants encounter in mixed group conversation
when using personal devices with Automatic Speech Recognition (ASR)
applications. We discuss problems, and describe accessibility
barriers in using these devices. We also describe best practices, as well as
lessons learned, and pitfalls to avoid in using personal devices in
conversation.
}

\section{Introduction}
Deaf or hard of hearing (DHH) people usually cannot understand speech unaided, and usually depend on additional support such as hearing aids or speech-to-text technology, especially in multi-speaker environments. Simple low-technology aids such as using paper and pen to write back and forth or to text back and forth can work, but are about 3-4 times slower than spoken or signed communication, and tends not to be effective for sustained communication.

\vspace{1em}
Deaf and hard of hearing use sign language interpreters and/or stenographers for access to auditory information.
They listen to the spoken information and translate into sign language. The stenographers also listen, and they type what they hear. Stenographers are intensely trained to use a special keyboard to be able to type different keystrokes to act as shortcuts for what they hear phonetically. The program then processes the phonetic sounds and matches it with its immense vocabulary database. If a mistake is made or if the program does not correctly identify a word, the stenographer has the ability to change it on the fly. Interpreters and stenographers are very good resources for providing access to deaf and hard of hearing individuals in a very timely manner. However, there are a few issues concerning the supply and dependability of these resources. If they become ill or are delayed and cannot come in, the deaf person would be stuck without them. Also, the cost of the service is very high, and is generally not affordable below the university level. The mental and physical work required to interpret or caption is heavy and is hard to maintain for a prolonged amount of time. Because of this multiple people are needed to hire so they can swap with each other and take turns. Small organizations and businesses would be very likely to not be able to afford professional interpreters or stenographers.

\vspace{1em}
This is where Automatic Speech Recognition comes in. The technology is very cheap, and ASR applications can be put on virtually any device. The only foreseeable physical issues with ASR technology is battery life, storage space, hardware quality, Internet connection, and portability. All of which have feasible solutions.

\vspace{1em}
When the participants use speech to text technology that is capable to keeping up with speech, they face challenges in following both the speaker's gestures and the extra visual of the speaker's speech-to-text. Deaf participants have to concentrate on managing competing tasks such as shifting attention between multiple visuals, compared with their hearing peers. They are left receiving incomplete information and trying to connect these segments together - all while searching for cues to know where to pay attention. As a result, even when provided with accurate real-time text through captioners, they receive only 50\%-80\% of the information, compared to 84\%-95\% for their hearing peers~\cite{marschark2008learning}. Similarly, hard of hearing participants have to make sense of reduced speech information through their hearing aids. 

\vspace{1em}
Deaf, hard of hearing and hearing people face diverse challenges with spoken language communication with each other in most conversational settings, especially in large-group and multiple-talker settings such as in classrooms and workplaces. They face a wide array of communication strategies and need considerable flexibility in accessible technology for upward mobility~\cite{foster_walter_undefined_1992}. Professional jobs in such areas as art, education, technology, and management demand more interaction and greater communication skills with hearing people than do nonprofessional occupations such as clerical, machine operations, printing, welding, food preparation, or janitorial.

\vspace{1em}
Job-related demands also make the workplace a more difficult communication situation for those who are deaf compared to those who are hard of hearing~\cite{boutin_2009}. Both groups, however, tend to experience less success in securing higher level jobs than their hearing peers and are limited by level of college degree~\cite{kellydeaf}. 

\vspace{1em}
For both deaf and hard-of-hearing people, communication on the job involves English about 80\% of the time, whether through writing, speech, or sign language with speech~\cite{kellydeaf}. Many deaf, hard of hearing and hearing participants have experimented with addressing these communication challenges through the use of ASR applications on their personal devices.

\vspace{1em}
Given the spoken-language communication requirements of the classroom and the workplace, we discuss how current ASR applications enhance access by deaf and hard-of-hearing individuals. We also examine how ASR applications enhance communication exchanges between deaf or hard-of-hearing persons and hearing persons in the classroom or workplace.

\section{Participants}
Deaf, hard of hearing and hearing speakers and listeners have different challenges and accessibility needs in mixed group conversation in most settings, including academic and workplace settings.

\subsection{Deaf Participants}
Deaf participants have challenges in both accessing and following spoken information and in conveying information efficiently and quickly to others in group settings. When deaf participants use ASR, they primarily use it to display and read the words of their communication partners. Sometimes they also use ASR to speak and convey information quickly to their communication partners especially if their speech clarity is sufficiently close enough to that of their hearing peers. The ability to display their own speech can aid communication by conveying both spoken and text information to their communication partners.

\vspace{1em}
For deaf participants, ASR is less accurate for them as compared to their hearing peers, because their speech intelligibility is lower on average. Venkatagiri et al.~\cite{venkatagiri2002speech} found that such systems could not recognize voice commands from subjects who had relatively low speech intelligibility and often those individuals were unable to correct their dictation errors. 

\vspace{1em}
Deaf speakers tend to have different segmental and prosodic factors, such as rate, pausing, voice volume, intonation, and stress, that may influence the overall performance of speech recognition applications and software. Jeyalakshmi et al.~\cite{jeyalakshmi2010deaf} found larger than normal variation in pitch and formants for deaf children, ages five to ten years, making such features unusable for recognition of deaf speech.

\subsection{Hard of Hearing Participants}
Hard of hearing people often do not face conversational challenges in quiet or one-to-one settings. They often face difficulties in multi-speaker or noisy settings, as they have difficulty in handling acoustics that interfere with the quality of the signal, extensive technical vocabulary, multiple information sources, and/or talkers with dialects or accents~\cite{Kushalnagar2014TACCESS}. Even if they use auditory assistive technologies that incorporate noise reduction algorithms that are capable of improving listening-alone performance, they cannot make up for the adverse effects of having to concentrate on following speech, and or dealing with competing tasks such as taking notes or shifting attention between multiple speakers or visuals \cite{doi:10.1044/hhdc25.1.24}. 

\subsection{Hearing Participants}
\begin{figure}
    \includegraphics[width=0.5\textwidth]{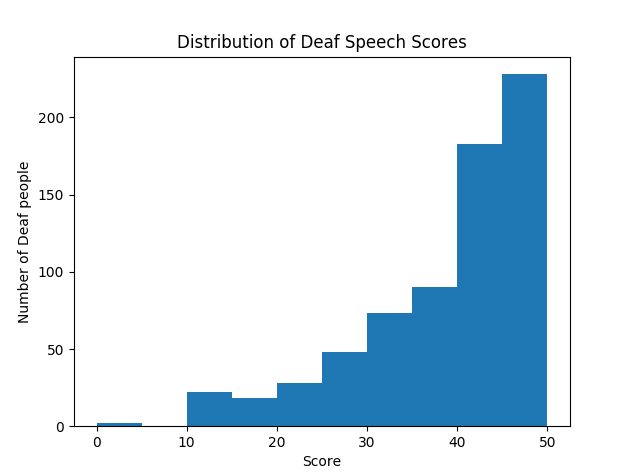}
    \caption{Distribution of Deaf Speech scores.}
    \label{fig:Statistics}
\end{figure}

While hearing participants do not typically have difficulty in speaking or listening to other hearing peers, they face difficulties in understanding deaf or hard of hearing speakers. This is due do their inability to adjust to the wide variance in speech production by deaf and hard of hearing speakers. In Figure \ref{fig:Statistics}, we show the distribution of about 650 deaf people as rated by professional speech pathologists at the National Technical Institute for the Deaf. Each of these ratings were based on a set of sentences called ``Clarke Sentences.'' In personal testing, scores from 1 to 3 were generally impossible to understand. A score from 3 to 4.5 was difficult to recognize, but it is still feasible for communication. Anything rated about a 4.5 and above should be understandable to most hearing people (with some effort). However amongst deaf people rated 5.0, ASR technology had a Word Error Rate of about 53\%, which is not useful enough for most applications.

\section{ASR Evolution}
Speech recognition is still a very difficult task for applications. For the last 50 years, researchers and inventors have iteratively implemented and improved applications that can understand speech, including conversation. 

\vspace{1em}
First-generation systems adopted a pattern-matching approach in which speech waveforms were matched with specific word waveforms. Because every talker produces a different waveform and a given talker often has different waveforms for the same word (e.g., if spoken excitedly, or with a cold), the first-generation systems could not be used outside controlled laboratory situations.

\vspace{1em}
Second-generation systems incorporated statistical prediction algorithms, mainly hidden Markov modeling (HMM). This approach was more robust to speech waveform variation, as it merged speech data from many talkers over time to build statistical models that captured the variation. As a result, speech recognition systems improved. In addition, faster computers and the ability to process dramatically more data through the cloud enabled many practical interactive speech systems. Most modern interactive phone systems are speech recognition systems that can answer simple questions. These systems, however, limit the domain of possible words in order to simplify analysis and increase accuracy. Unfortunately, for unrestricted domains, even the best second-generation speech recognition systems still had word error rates of 20-25\% with arbitrary speech~\cite{Kushalnagar2014TACCESS}.

\vspace{1em}
Third generation systems now have the potential to assist deaf and hard-of-hearing users. These systems incorporate learning algorithms, which try to emulate human brain behavior. Driven by companies such as Apple, Microsoft, and Google, with intended applications in wearables, cars, robotics, and machines, these systems seem to handle human speech variation better. In fact, in many commercial systems, word (or character) error rates for tasks such as mobile short message dictation and voice search are far below 10\%. Some companies are even aiming at reducing the sentence error rate to below 10\%~\cite{yu2014automatic}. 

\vspace{1em}
However, as constraints are relaxed, allowing longer utterances, technical jargon, and environmental noise, speech recognition systems still have word and character error rates of about 20\%, as reported by Yu et al.~\cite{yu2014automatic}, under conditions such as:

\begin{itemize}
\item Far field microphones (e.g., when the microphone is in the background, as in a living room, meeting room, or audiovisual recording made in the field)
\item Noise (e.g., when loud music is captured by the microphone)
\item Speech produced with an accent
\item Multi-talker speech or side conversation (as in a meeting or with multiparty chatting)
\item Spontaneous speech that is not fluent, with variable speeds, or with emotion.
\end{itemize}



\section{ASR for Conversational Use}
Major obstacles that limit ASR use in group conversation for deaf and hard of hearing includes text accuracy and lag time. Additionally, ASR accuracy for any application can be affected by other variables including auditory factors such as speech fidelity, ambient noise and microphone quality, and computing factors such as available computing power, speech recognition engine and associated statistical models. 

\vspace{1em}

\subsection{Word Accuracy}
The text accuracy for ASR needs to be sufficiently high to be usable and beneficial. Wald et al.~\cite{wald2005using} found that learning improved for students with disabilities when using such technologies, but only if the text was at least 85\% accurate. Others have suggested that a speech recognition transcript must be at least 90\% accurate to be useful in the classroom~\cite{kheir2007inclusion} and that an even greater accuracy of 98\% or higher may be needed to maintain the meaning and intent of the message~\cite{stuckless1999recognition}, actual accuracy rates are most often too low for use by deaf and hard-of-hearing students in typical higher education settings~\cite{kheir2007inclusion}.

\subsection{Transcription Speed}
Similarly, the lag time for ASR needs to be short enough to be usable. Lag time for ASR becomes worse as the amount of data to be analyzed increases and causes processing delays. It has been shown that students cannot effectively participate in discussions or dialogues if lag time is more than 5 seconds~\cite{wald2005using}. In a study of college students who viewed transcriptions of a lecture, both deaf and hearing participants commented that captions created by automatic speech recognition software were choppy and had too much latency, making it hard to follow, compared to captioning by a stenographer or a crowd-captioning process~\cite{Kushalnagar2014TACCESS}.

\section{User Experience}
To investigate the capabilities of current ASR applications, from Fall 2016 through Summer 2017, the authors used a variety of applications on personal devices in everyday, real-world settings. The purpose of using the speech recognition applications was to facilitate face-to-face spoken language interactions by providing a visible text representation of speech in the following contexts: 1) classroom communication, 2) conversation, 3) job interviews, and 4) speech production practice in which the text displayed by an app was used as an indicator of the intelligibility. 

\vspace{1em}
Application software, also known as an ``application'' or ``app'', enables an user to carry out specific tasks on the computer. The authors used and evaluated communication apps described in the list below. These apps were chosen because they were available at no additional cost, other than device charges and service plans, and had been rated at least 3.5 out of 5 for user satisfaction in evaluations published by the developers. 

\vspace{1em}
The developers of each product described them as follows:
DEAFCOM by askjerry Communications "is a very simple application that will convert speech into readable text. For a non-hearing or hard-of-hearing person, the application will allow faster communication with deaf persons. For deaf users, the software can assist in faster communication and may also be used as a useful tool when practicing your speech"~\footnote{\url{play.google.com/store/apps/details?id=defcom.v1}}.

\vspace{1em}
Dragon Dictation by Nuance Communications is described as "an easy-to-use voice recognition application powered by Dragon Naturally Speaking that allows you to easily speak and instantly see your text or email messages. In fact, it's up to five (5) times faster than typing on the keyboard"~\footnote{\url{www.dragonmobileapps.com/android/}}. 

\vspace{1em}
Siri by Apple Inc., part of iOS, is described in these words: ``Talk to Siri as you would to a friend and it can help you get things done, like sending messages, placing calls, or making dinner reservations. You can ask Siri to show you the Orion constellation or to flip a coin. Siri works hands-free, so you can ask it to show you the best route home and what your ETA is while driving. And it's connected to the world, working with Wikipedia, Yelp, Rotten Tomatoes, Shazam, and other online services to get you even more answers. The more you use Siri, the more you'll realize how great it is. And just how much it can do for you''~\footnote{www.apple.com/ios/siri/}.
 
\vspace{1em}
Virtual Voice by Gareth Hannaway Communications is ``designed to use the text to speech (TTS) and the speech recognition features of your Android device. It was created with deaf and/or mute people in mind, so they can communicate with others without the need for sign language or lip reading''~\footnote{\url{play.google.com/store/apps/details?id=appinventor.ai\_Gareth\_Hannaway\_420.VirtualVoice}}.

\vspace{1em}
Ava is is an eponymous product, which is described as follows: ``Ava shows you who says what. Ava shows you what people say, in less than a second. Easy communication is only a tap away.''~\footnote{\url{https://www.ava.me/}}

\vspace{1em}
The Google Assistant is from Google Inc. ``Meet your Google Assistant. Ask it questions. Tell it to do things. It's your own personal Google, always ready to help. Ready to help, wherever you are. Your one Assistant extends to help you across devices, like Google Home, your phone, and more. Discover what your Assistant can do. Learn more about how you can get help from your Assistant. Works with your favorite stuff, too. Shuffle your favorite playlist, dim your Philips Hue lights with just your voice, or ask your Assistant on Google Home to stream Netflix to your TV with Chromecast. Discover more services and smart devices that work with your Google Assistant.''~\footnote{\url{https://assistant.google.com/}}

\vspace{1em}
Alexa is owned by Amazon.com, Inc. ``Using Alexa is as simple as asking a question. Just ask to play music, read the news, control your smart home, tell a joke, and more-Alexa will respond instantly. Whether you are at home or on the go, Alexa is designed to make your life easier by letting you voice-control your world. Alexa lives in the cloud so it's always getting smarter, and updates are delivered automatically. The more you talk to Alexa, the more it adapts to your speech patterns, vocabulary, and personal preferences. Alexa comes included with Echo and other Alexa devices.''~\footnote{\url{https://www.amazon.com/Amazon-Echo-And-Alexa-Devices/b?ie=UTF8&node=9818047011}}

\vspace{1em}
The authors downloaded the chosen apps to their personal iPhone or Android device and evaluated use in both quiet one-to-one settings and in noisy, open areas with multiple speakers.  

\begin{figure}
    \includegraphics[width=0.5\textwidth]{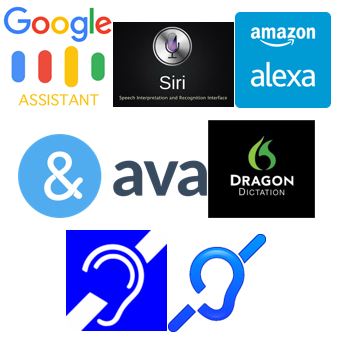}
    \caption{Logos of various ASR platforms used}
    \label{fig:Logos}
\end{figure}

\vspace{1em}
Specifically, the aim was to assess how well these technologies could facilitate: 1) classroom communication, 2) informal conversations, and 3) speech production practice. The technologies selected for evaluation were DEAFCOM, Dragon Dictation, Siri, Virtual Voice, Ava, Google Assistant, and Amazon Alexa. The logos for these products can be seen in Figure \ref{fig:Logos}.

\vspace{1em}
All apps were both accurate and had minimal lag in specific quiet one-to-one settings with some hearing peers, and the authors found that they could use them extensively in these ideal settings. For many other quiet one-to-one settings, the apps were less accurate and had more lag, probably because the speaker had an accent or had some background noise that was not noticed during the conversation.

\vspace{1em}
The performance of all apps did not serve either our needs or our peer's needs in noisy, open areas with multiple speakers. Performance was disappointing when there was any level of noise in the environment and when the talker had an accent. Excessive lag time in displaying a talker's speech also was a critical factor that resulted in unacceptable communication problems. There is much work yet to be done to make these technologies useful enough to put away pencil and paper.

Our hearing peers who had concerns about communicating with deaf or hard of hearing individuals felt more at ease when they were able to use the apps, even in noisy or multi-speaker settings. Since these devices are ubiquitous, the mere knowledge that they could rely on these apps seemed to put them at ease. 

\vspace{1em}
The speech recognition apps tested often inserted random text, especially in noisy settings. Collection of more training data for far field microphones might lead to improved performance. Additionally, new paradigms in acoustic modeling will be needed. Future speech recognition systems will need to learn the key speech characteristics from the training set and, to be acceptable to users, will need to generalize well to unseen speakers and accents, even in noisy conditions~\cite{yu2014automatic}

\vspace{1em}
Most apps did not recognize the different prosody, pitch, and articulations of deaf speakers. Specific apps that included optimization for deaf or hard of hearing speakers had less lag time, but still had high error rates. For purposes of facilitating speech intelligibility, most of our deaf and hard-of-hearing evaluators, regardless of their mode of communication, did not find the speech recognition apps to be usable. They were quite disappointed to find that they uniformly failed to recognize their speech and they tended to attribute this failure to differences in their own articulation, pitch, and prosody. The alternative of switching to a text-to-speech or text-to-text function resulted in significant slowing in conversational interactions. 

\vspace{1em}
In most situations, the settings were not ideal, and the level of accuracy and degree of latency characteristic of the apps were not adequate to enhance speech reception in face-to-face interaction. We experienced a disruptive display of text that did not match in time what the others were saying. More needs to be done so that the text is synchronous with the speech. When the authors ignored the speech or did not speechread, it was useful to have access merely to the overall context of the message via keywords that the apps could display. 

\vspace{1em}
Our use of ASR apps drew both deaf and hearing people into a collective learning experience:

\begin{quote}
It's telling me that at least I'm not the only person that might have a problem understanding. Like, I know that sometimes when you've got a disability you feel like you're the only one ... I just don't want it to benefit us. I'd like to see it work for everybody
\end{quote}

\begin{quote}
We are not the only ones with problems understanding .... That we're not the only ones. There's non-disabled people out there who are having the same problems ... you feel equal.
\end{quote}

\section{Lessons Learned}
When using ASR technology, deaf, hard of hearing and hearing people report different ways of interacting with the technology. The following lessons were learned from daily experience with the technology.

\begin{enumerate}
\item Hearing people often use the technology when their hands are occupied as a way to do basic tasks such as looking up simple queries on the web. They do not generally consider speech to text translations to be particularly important.
\item Deaf people most often reported that their experience with the technology for their own use was restricted to messing around with the technology in order to see what the technology might interpret from their speech. Some report that ASR is starting to work better for them now, to the point where some deaf speech can be comfortably understandable.
\item One deaf person reported that she commonly would use Amazon Alexa for its intended purposes and said that she was satisfied with its use. Generally most deaf people we met have not had this level of success with the technology yet.
\item So far, only a small portion of the hearing population seems to often use the voice recognition feature of their phone. Several reported that the technology does not work well enough for them or that they felt uncomfortable using the technology in a public setting. This led to them not using the technology in a private setting either.
\item ASR services generally use Internet connections to send audio data to their service. Phones have various connection strengths and reliability. Some have difficulty with wireless networks and connections, so performing the actual analysis can take erratic amounts of time. Also, using ASR while not on Wi-Fi may result in additional data, battery, space usage and costs.
\item When using ASR to communicate with another person, the ability to change the text on the fly has been limited and/or not feasible. Many people often mention this when discussing why they do not use ASR as much as texting.
\item Interaction with ASR devices such as Alexa and the Google Home has been limited since deaf people do not have access to the verbal responses from said devices after commands are spoken to it. The two most popular personal assistants for smart phones, Google Assistant and Siri display their commands, but sometimes only voice their responses. It was not until a couple of months ago that both personal assistants have introduced text inputs in addition to voice inputs. This makes them more accessible for DHH people, however this removes much of the appeal of a personal assistant.
\item A few deaf people have reported that they experience obstacles when they try to converse with a hearing person who have never used ASR technology before. They reported that this has been mostly a result of the user interface design of the ASR app itself and the fact that they were not intuitive to use.
\item Users said that it would be ideal if ASR systems could tell the user if repetition of a specific word was needed rather than the whole word. Using a system like this might increase the interaction of users with the device. 
\item Mild accents do not affect ASR much anymore due to advances in the technology. Heavy accents or unique accents still confuse the software too much to be usable on a daily basis.
\end{enumerate}


\vspace{1em}
While improvements in speech recognition are being reported with third generation processing strategies in laboratory settings~\cite{hannun_2015}, deaf users are still noting performance issues in the real world, especially in the critical settings of the classroom and on the job. Improvements are needed, particularly to control noise and side-talk interference, perhaps with better noise canceling algorithms, more advanced microphone array techniques, and through use of a lapel mic, Bluetooth streaming, and/or Wi-Fi Direct, and to ultimately convince deaf and hard-of-hearing persons that speech recognition technologies are better than pencil and paper when trying to communicate to a person with typical hearing when it counts in the classroom or on the job.


\vspace{1em}
These apps can produce text or spoken output for users with speech that is difficult to understand. However, they does not mean they will actually use it. They may prefer to use more reliable, easier or less conspicuous low-tech solutions such as paper and pen.

\vspace{1em}
The idea today is that everyone should make a small effort to make the conversation work. Of course a minimal effort, but the burden should not be on the deaf person alone. Everyone works together.

\vspace{1em}
Many deaf speakers have low volume, and directional microphones can make a significant difference. It would be helpful to include external microphone support for microphones that can be either plugged in or connected with Bluetooth. This could be a lapel type of mic that is clipped on the person, or a more central hardware piece with multiple microphones. These could indeed do both directional beam forming or omni directional capturing. Most applications need fairly loud speech samples for optimal accuracy. Deaf people typically cannot monitor their own speech volume, and it is important to provide users good feedback about good placement of the phone and associated speech volume.

\section{Conclusions}

Although many people use ASR systems such as Siri or Google for recreational use and every once in a while to send a text, they are not comfortable using these systems for sustained conversational use, as the systems have higher than tolerable error rates, especially in less than perfect settings. Whenever there are errors, the errors are time consuming to fix and the interfaces are not customizable.

\vspace{1em}
ASR technology has been around for over 50 years and has been constantly updated and improved by many researchers and companies. However, ASR technology has been focused on the hearing population and therefore it is difficult for deaf and hard of hearing individuals, even those who use voice on a regular basis, to be fully comfortable with ASR. As found in one of our studies, there is a big variance in WER for those whom have voices that are understandable by a hearing person. This means that even though a deaf person might seem to speak well, ASR still won't always have good results. On the other hand, The variance was very small for those whose voices scored low on the intelligibility scale. This means that for these people, they have almost no hope using ASR technology with their own voices. This, along with several other factors, shows that deaf people will have various experiences with ASR technology, and that ASR is not stable or reliable to use with Deaf people at this time.

\vspace{1em}
Putting aside the problems facing deaf people using their own voices with ASR, there are still user interface accessibility issues. For example, if it was practical for ASR technology to be used in conversations between a deaf person and a hearing person, the ASR interface should be intuitive and easy to use. We have seen that current technology is still not always convenient. People have frustrations with lack of Internet connections and battery and space usage, so it should be an objective to make ASR apps efficient. People who have little or no experience using ASR technology may not know how to interact with the app. Thus, clear, intuitive user interfaces need to be added to the ASR apps so that the conversation will have better flow and comfort. Apps are typically limited and have a certain amount they will transcribe in a time period. It has been observed that some apps require purchases for more usage. This is disadvantageous because Deaf people should not have to work harder and give up more to have equal access to information.

\vspace{1em}
When ASR technology is improved upon, it is critical to take into consideration all of the factors and perspectives. It would be very beneficial to have an experiment along with a survey and to recruit people with different backgrounds. This way, the developers and researchers will be able to work with accessibility in mind and have their products benefit a wide spectrum of people. This process applies to all technology, especially those that become widely available, commercial products.

\bibliographystyle{unsrt}
\bibliography{ux}

\begin{thebibliography}{10}

\bibitem{marschark2008learning}
Marc Marschark, Patricia Sapere, Carol Convertino, and Jeff Pelz.
\newblock Learning via direct and mediated instruction by deaf students.
\newblock {\em Journal of deaf studies and deaf education}, 13(4):546--561,
  2008.

\bibitem{foster_walter_undefined_1992}
Susan~Bannerman Foster, Gerard~G. Walter, and undefined~undefined undefined.
\newblock {\em Deaf students in postsecondary education}.
\newblock Routledge, 1992.

\bibitem{boutin_2009}
Daniel~L Boutin.
\newblock Professional jobs and hearing loss: A comparison of deaf and hard of
  hearing consumers.
\newblock {\em Journal of Rehabilitation}, 75(1):36--40, Mar 2009.

\bibitem{kellydeaf}
Ronald~R Kelly.
\newblock Deaf workers: Educated and employed, but limited in career growth.

\bibitem{venkatagiri2002speech}
HS~Venkatagiri.
\newblock Speech recognition technology applications in communication
  disorders.
\newblock {\em American Journal of Speech-Language Pathology}, 11(4):323--332,
  2002.

\bibitem{jeyalakshmi2010deaf}
C~Jeyalakshmi, V~Krishnamurthi, and A~Revathy.
\newblock Deaf speech assessment using digital processing techniques.
\newblock {\em Signal \& Image Processing: An International Journal (SIPIJ)},
  1(1):14--25, 2010.

\bibitem{Kushalnagar2014TACCESS}
Raja.~S. Kushalnagar, Walter~S. Lasecki, and Jeffrey~P. Bigham.
\newblock {Accessibility Evaluation of Classroom Captions}.
\newblock {\em ACM Transactions on Accessible Computing}, 5(3):1--24, 2014.

\bibitem{doi:10.1044/hhdc25.1.24}
Karen~L. Anderson.
\newblock Access is the issue, not hearing loss: New policy clarification
  requires schools to ensure effective communication access.
\newblock {\em SIG 9 Perspectives on Hearing and Hearing Disorders in
  Childhood}, 25(1):24--36, 2015.

\bibitem{yu2014automatic}
Dong Yu and Li~Deng.
\newblock {\em Automatic speech recognition: A deep learning approach}.
\newblock Springer, 2014.

\bibitem{wald2005using}
Mike Wald.
\newblock Using automatic speech recognition to enhance education for all
  students: Turning a vision into reality.
\newblock In {\em Frontiers in Education, 2005. FIE'05. Proceedings 35th Annual
  Conference}, pages S3G--S3G. IEEE, 2005.

\bibitem{kheir2007inclusion}
Richard Kheir and Thomas Way.
\newblock Inclusion of deaf students in computer science classes using
  real-time speech transcription.
\newblock {\em ACM Sigcse Bulletin}, 39(3):261--265, 2007.

\bibitem{stuckless1999recognition}
Ross Stuckless.
\newblock Recognition means more than just getting the words right: Beyond
  accuracy to readability.
\newblock {\em Speech Technology}, 1:30--35, 1999.

\bibitem{hannun_2015}
A.~Hannun.
\newblock Deep speech: Lessons from deep learning, 2015.

\end{thebibliography}

\end{document}